\documentclass{PoS}

\newcommand{\farcs}{\mbox{\ensuremath{\!\!^{\prime\prime}}}}
\newcommand{\igra}{\mbox{IGR~J16465$-$4507}}
\newcommand{\igrb}{\mbox{IGR~J18483$-$0311}}
\newcommand{\igrc}{\mbox{IGR~J16418$-$4532}}
\newcommand{\be}{\begin{enumerate}}

\title{First results of X-shooter observations of IGR sources}

\ShortTitle{First results of X-shooter observations of IGR sources}

\author{\speaker{P. Goldoni}\thanks{Based on observations collected at
  the European Organization for Astronomical Research in the Southern
  Emisphere, Chile under Program P089.D-0056(A)}\\
        APC, Univ. Paris Diderot, CNRS/IN2P3, CEA/IRFU, Obs. de Paris,
        Sorbonne Paris Cit\'e, France\\
        E-mail: \email{goldoni@apc.univ-paris7.fr}}
\author{S. Chaty\\
        Laboratoire AIM (UMR-E 9005 CEA/DSM-CNRS-Universit\'e Paris Diderot),
Irfu/Service d'Astrophysique, CEA-Saclay and Institut Universitaire de
France, 103, bd Saint Michel, Paris, France}
\author{A. Goldwurm\\
        APC, Univ. Paris Diderot, CNRS/IN2P3, CEA/IRFU, Obs. de Paris,
        Sorbonne Paris Cit\'e, France}
\author{A. Coleiro\\
        Laboratoire AIM (UMR-E 9005 CEA/DSM-CNRS-Universit\'e Paris Diderot),
Irfu/Service d'Astrophysique, CEA-Saclay}

\abstract{
X-shooter is a second generation ESO-VLT instrument 
that had its first light in October 2009. It is a single object
medium-resolution spectrograph whose main feature
is the capability of covering simultaneously in a single
observation the range from 3000 to 24000 Angstrom.
This unique capability is very well suited to investigate
the complex spectra of the optical counterparts of X/gamma
ray sources which usually display signatures of different
components in emission and absorption.
In July 2012 we observed with X-shooter a small sample of
counterparts of bright IGR sources in order to better determine
their physical properties. We present the first results of these
observations.
}
\FullConference{'An INTEGRAL view of the high-energy sky (the first 10 years)'
9th INTEGRAL Workshop and celebration of the 10th anniversary of the launch,\\
		October 15-19, 2012\\
		Bibliotheque Nationale de France, Paris, France}

\begin{document}

\section{Introduction}

 INTEGRAL has discovered several bright sources in the Galactic plane, they are usually divided in two groups: obscured sources   
and SFXTs \cite{cha08}. Their study is difficult due to their distance (> 1-15 kpc), absorption (line-of-sight and intrinsic)
and the crowded fields in which they are found. Since their first discovery, several of them have been identified as High Mass X-ray Binaries.
The vast majority of them are quite bright in infrared and faint in the visible range due to extremely high absorption (A(V) almost
always greater than 10). Therefore the spectroscopic observations have almost always been performed in infrared and for
some of these sources, no visible spectrum is available. This is a problem because the best diagnostics of the 
properties of the primary star such as rotation, metallicity and mass
loss lie in the visible and ultraviolet domains (see e.g.
\cite{mar10} ). For the sources that do have visible spectra, they are not simultaneous with the infrared ones which may make the
comparison unreliable due to the intrinsic variability of the primary star at least concerning the emission features.
 
 The X-shooter spectrograph \cite{ver11} with its exceptional wavelength range and its high sensitivity is very well placed to investigate
on these matters allowing to monitor the known NIR spectrum while exploring the VIS and UVB spectrum whenever possible.
 In principle the comparison of the simultaneous visible and infrared diagnostics should allow to constrain strongly the properties 
of the primary star. Moreover the line of sight absorption can be studied thoroughly by measuring the total reddening with the SED
and measuring the equivalent width and, for identified lines, the Doppler shifts of ISM absorption features.
We asked for 2.5 nights of French GTO time to observe a sample of eleven IGR sources in order to perform a first test of X-shooter
 capabilities on these sources. We aimed at observing all the sources of the sample at least twice to check for variable features.
 Unfortunately, due to bad weather, we could observe only one night and therefore only seven sources could be observed once. 
Here we report on the first results of these observations for three of the sources in our sample.

\section{Observations and Data Reduction}

  The main observations consisted of four different pointings of 300 s
each taken using the nodding along the slit technique with an
offset of 5 arcsec between pointings in a standard ABBA sequence.
In order to avoid saturation, all the exposures on the NIR arm and some
of those in the VIS arm were split in shorter integrations.
All of these pointings were performed using narrow slits (0.5 \farcs~for
the UVB arm, 0.7 \farcs~for the VIS arm and 0.6 \farcs~for the NIR
arm) (see table \ref{obssum} for details).
Each source was also briefly observed with a wide (5.0 \farcs)
slit to estimate the slit losses. A telluric A0V star was observed
before each source and a flux standard was observed in
the beginning of the night.

The spectra were reduced using the X-shooter pipeline
\cite{gol06,mod10}. We used the pipeline to build
calibration files (master bias, wavelength
solution...) from daytime calibration observations. Afterwards we
performed bias and dark subtraction followed by division
by a master flat field.  The orders were then extracted, rectified in
wavelength-slit space and then shifted and added to subtract the
sky contribution. The resulting 2D order spectra were then merged
to produce the final 2D spectrum and the 1D spectrum was extracted
using a PSF-weighing scheme in IDL.
The 1D spectra were subsequently normalized. Flux calibrations and
telluric corrections have not been performed yet.

\begin{table}[h!]
\begin{center}
\caption{ Summary of the narrow slit observations for the three sources presented
  here. Four independent pointings of 300s each were performed in a standard
nodding ABBA sequence. Each pointing in the NIR arm was split
in 20 independent exposures each one with up to 6 independent
integrations that are averaged by the detector. The exposure time is
given as N(exposure) x N(integrations) x (Detector Integration Time or
DIT).}
\label{obssum}
\begin{tabular}{cccc}

\hline

Source Name & Coordinates & Exptime (s) &Slit Width (\farcs)\\
                     &  (J2000)       &     UVB, ~VIS, ~NIR &
                    UVB, ~VIS, ~NIR \\
\hline
\igra    & 16 46 35 -45 07 04 & 4 x 300,~4 x (2x150), ~20 x (2x30)  &
0.5 \farcs, ~0.7 \farcs, ~0.6 \farcs \\
\hline
\igrb    & 18 48 17 -03 10 15 & 4 x 300, ~4 x 300, ~20 x (6x10) &
0.5 \farcs, ~0.7 \farcs, ~0.6 \farcs \\
\hline
\igrc   & 16 41 51   -45 32 25  & 4 x 300, ~4 x 300, ~20 x (2x30) &
0.5 \farcs, ~0.7 \farcs, ~0.6 \farcs \\
\hline
\end{tabular}
\end{center}
\end{table}

\section{\igra}

This source has been classified as a supergiant of spectral type B0-B1 (\cite{neg06},\cite{cha08}) or O9.5 \cite{nes08}.  
Spectroscopic observations have been performed from the U (3900 \AA)
to the K band.  We clearly detect \igra~in the UVB arm starting from
about 3500 \AA. The UVB spectrum looks similar to the one shown by
\cite{neg06} . The possible presence of the HeII 4686 line \AA~suggests
a B0.5-B1 spectral type \cite{eva04}. The width of thin lines such as e.g.
SiIII 4552  \AA~suggests fast rotation with vsini greater than 200 km/s. The
high rotation velocity implies that we're observing the source at high
inclination. The VIS spectrum is also quite bright showing an intense H$\alpha$
feature  (not shown) and several He I features. 

\begin{figure}[h!]
\begin{center}
\begin{tabular}{ccc}

\includegraphics[width=.4\textwidth,height=5truecm]{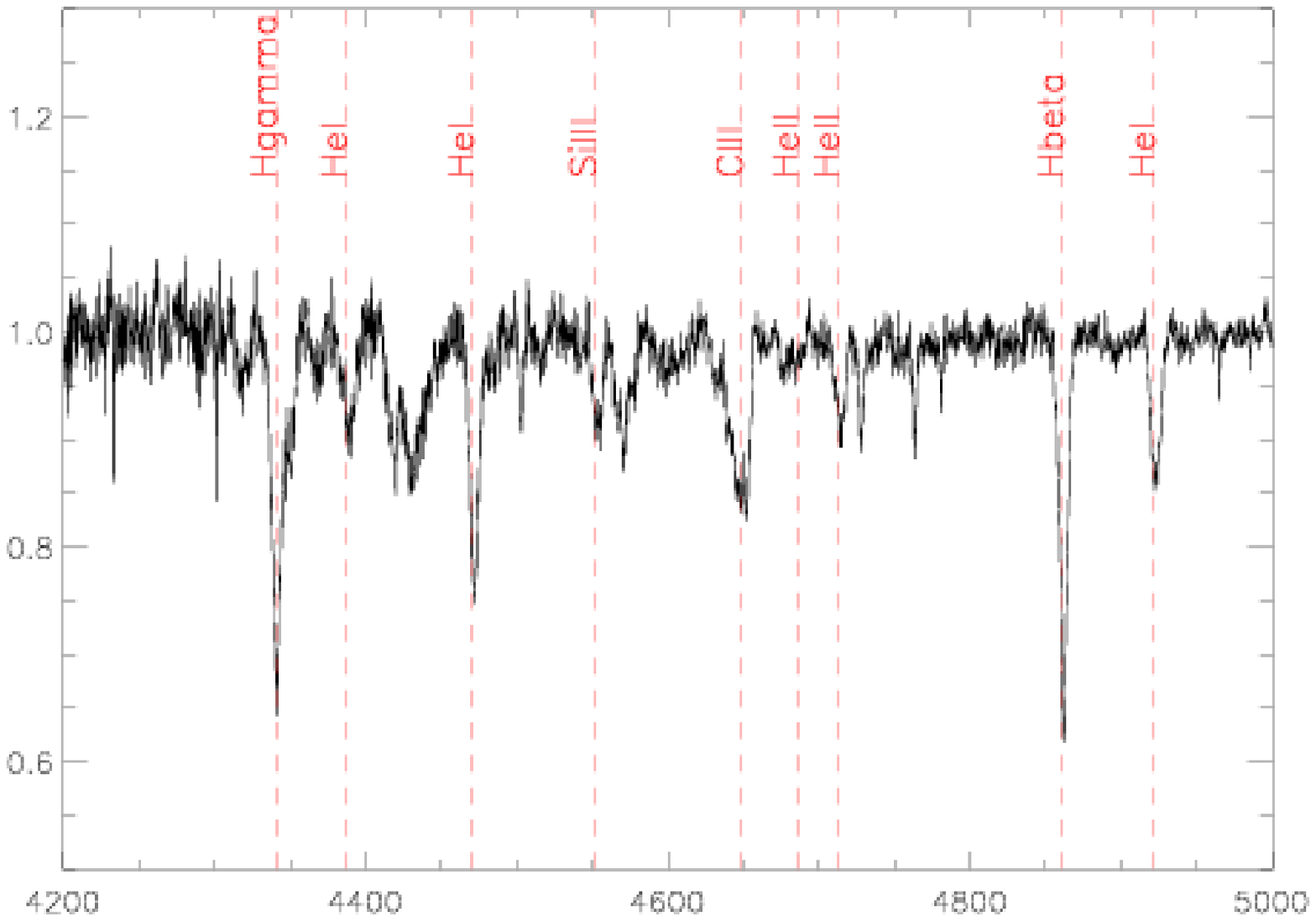} &
\includegraphics[width=.08\textwidth,height=5truecm]{white.ps} &
\includegraphics[width=.4\textwidth,height=5truecm]{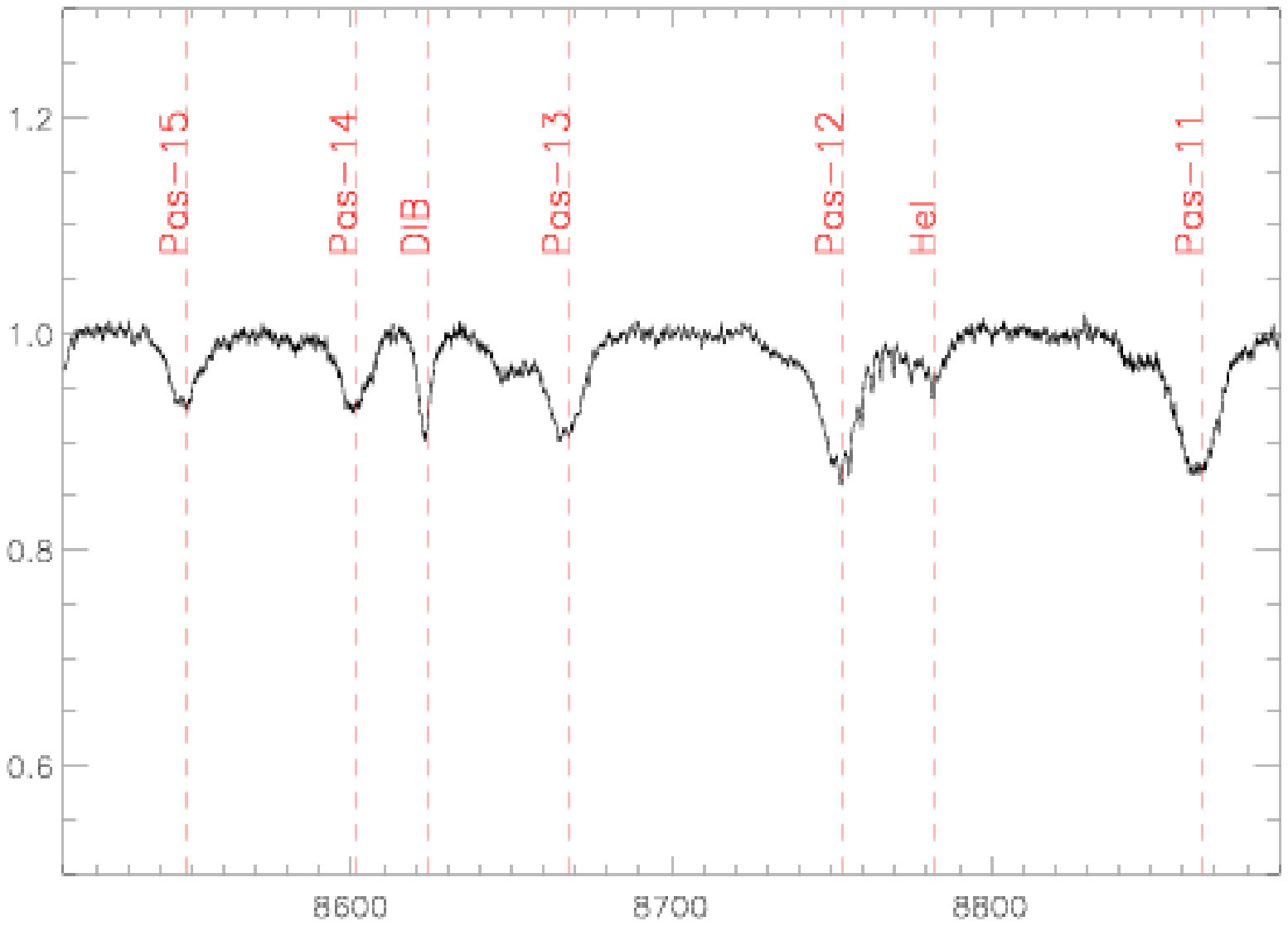} \\

\end{tabular}

\end{center}
\caption{Left: Section of the UVB spectrum of \igra, the
  possible presence of a weak HeII 4686 \AA~line suggests a
spectral type B0.5-B1 \cite{neg06} as for an O star this line would be stronger.
 Right: Section of the I band spectrum of \igra, containing several
Paschen lines and the  8620 \AA~DIB with equivalent width $\sim$
0.5 \AA.}
\label{fig1}
\end{figure}

 In figure \ref{fig1} we show the first I band spectrum of \igra,
preliminary comparisons with the I band spectra of Westerlund 1
supergiant stars \cite{neg10} also suggest a B0.5-B1 spectral type.
More quantitatively, we estimated the spectral type with the
properties of the Paschen lines using the approach of \cite{car03}. We
measured the equivalent width of the Paschen-11 and Paschen-17 lines
obtaining $\sim$2.4 and $\sim$0.6 \AA~respectively. These values,
once compared to Figure 7 and 8 (last panel) of \cite{car03} 
again point toward a spectral type B0.5- B1 with Teff $\sim$25000 K
and logg $\sim$3.

 We also remark that the 8620 \AA ~DIB is clearly detected with
EW$\sim$0.5 \AA. This is expected considering the source 
absorption. The equivalent width of this DIB has been proposed as a reddening
indicator by \cite{mun08}. Using their formula we can estimate
E(B-V) $\sim$1.36 which implies A(V) $\sim$4.2. This value is outside by
about one magnitude of the 90\% confidence range estimated by
\cite{rah08_1}, however using R(V)=4, we obtain A(V)=5.44, well inside their
confidence range. This may suggest a non standard extinction
law in this direction.

\section{\igrb}

\begin{figure}

\begin{tabular}{ccc}

\includegraphics[width=.4\textwidth,height=5truecm]{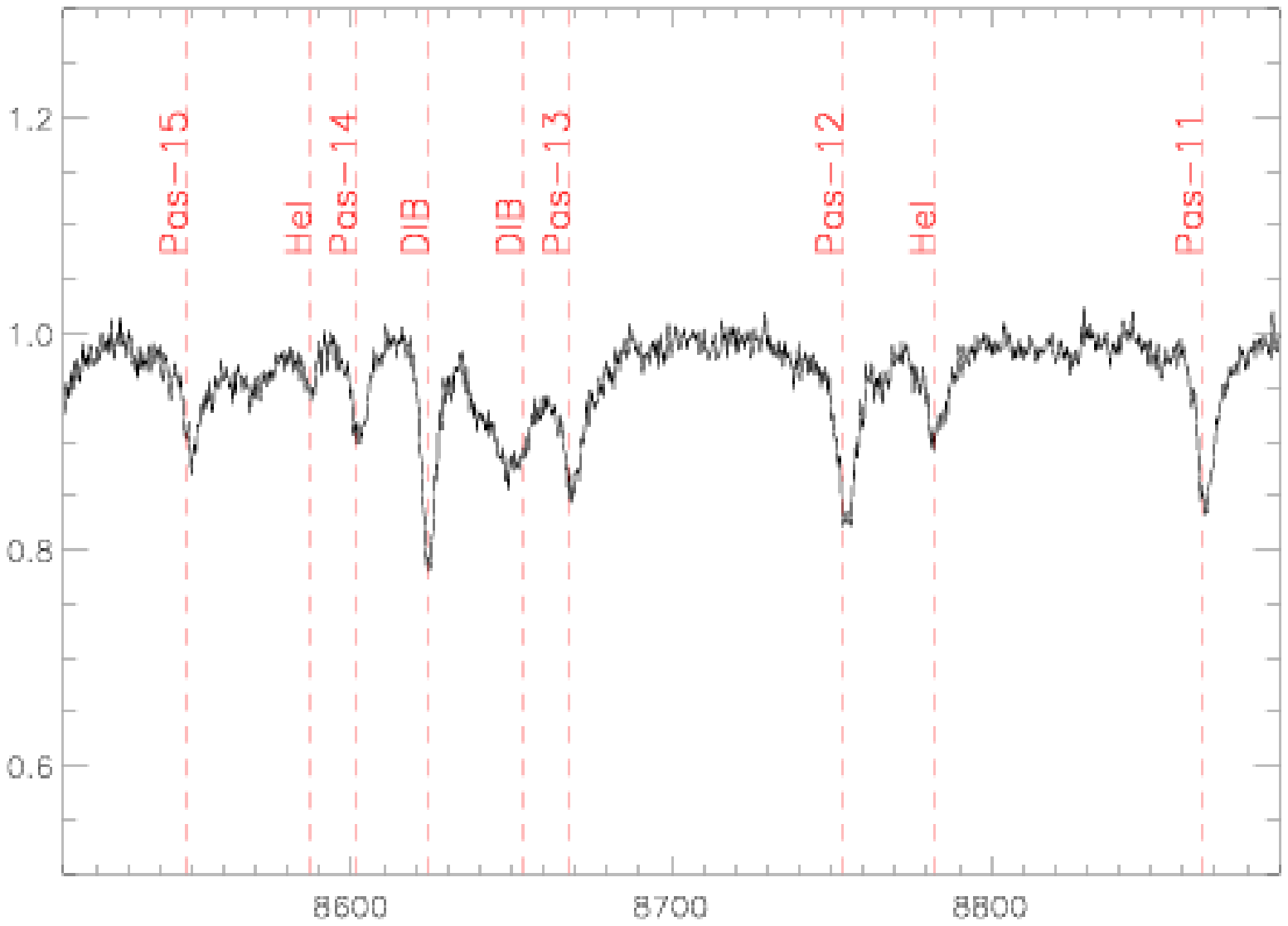} &
\includegraphics[width=.08\textwidth,height=5truecm]{white.ps} &
\includegraphics[width=.4\textwidth,height=5truecm]{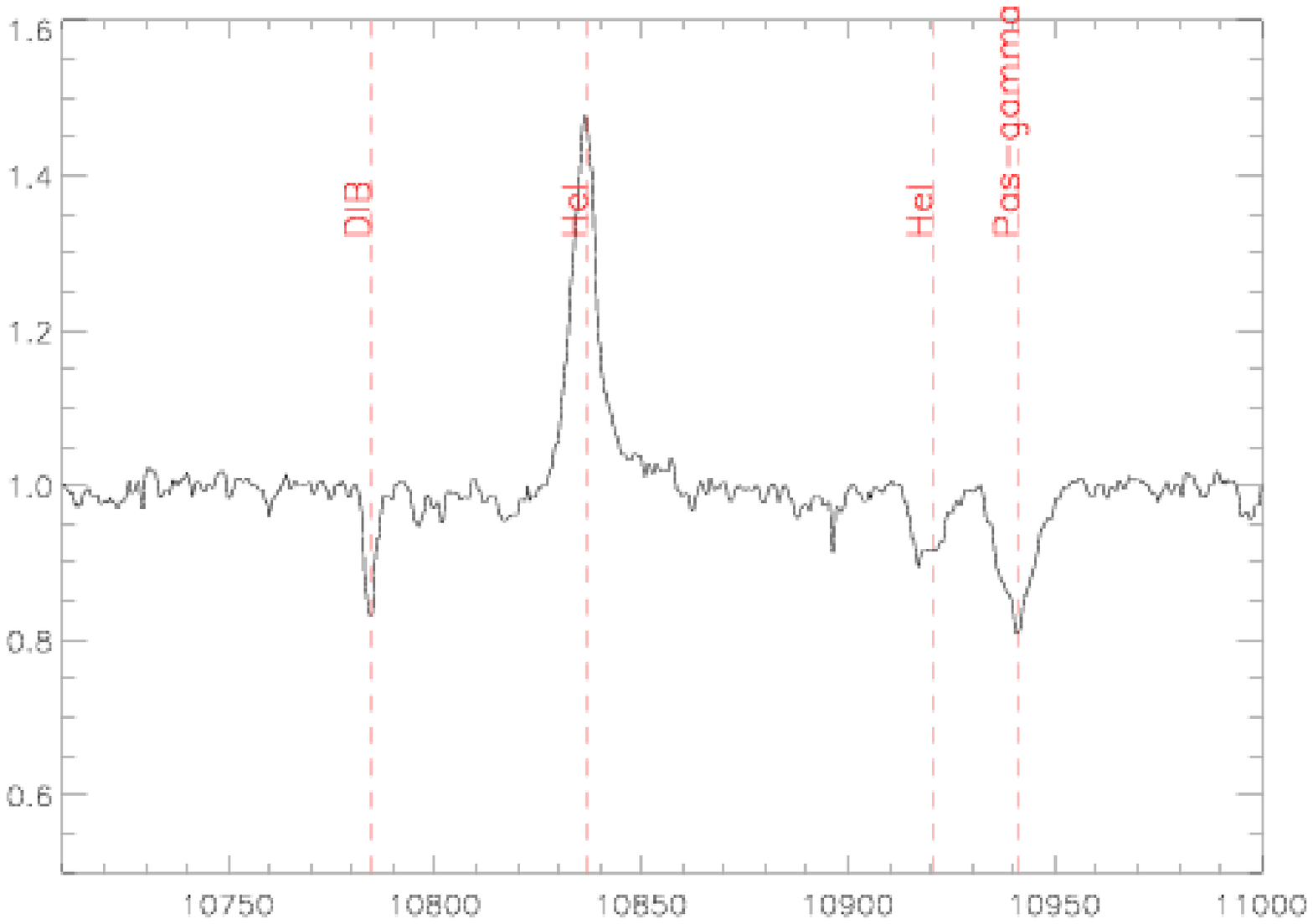} \\

\end{tabular}

\caption{Left: section of the I band spectrum of \igrb~containing several Paschen lines
and the DIBs at 8620 \AA~and 8643 \AA.
The equivalent width of the 8620 \AA~ line is $\sim$1.2 \AA.
Right: Section of the J band spectrum of \igrb~containing the HeI 10830 \AA~line in emission,
the HeI 10920 \AA~and Brackett $\gamma$ lines in absorption. The HeI 
10830 \AA~ emission line has a slightly asymmetric profile which may be
of P Cygni type however a nore careful analysis is required. The
feature at 10780 \AA~proposed as a possible DIB by \cite{gro07} is also present.}
\label{fig2}
\end{figure}

\igrb~has been classified as a B0.5 supergiant with NIR spectroscopic
observations \cite{rah08} and it has also been observed non simultaneously
in optical \cite{mas08}, \cite{tor10}.
We detected \igrb~only starting from about 6000 \AA, a weak
H$\alpha$ line is visible, similar to \cite{mas08}. From the
I band spectrum in Figure \ref{fig2} we measured the equivalent width of the
Paschen-11 and Paschen-17 lines obtaining respectively 2.1 \AA~and 0.6 \AA~
thus validating the spectral classification by \cite{rah08}.
We detect the 8620 \AA~DIB with EW $\sim$1.2 \AA~and
therefore E(B-V) $\sim$3.26 which is in good agreement
with published photometry \cite{rah08}. From it we obtain A(V)=10.2 for
R(V)=3.1, smaller than 15.7 as in \cite{rah08}. Again a non standard
extinction law with R(V)$\sim$4.8 could explain this discrepancy.

 In the NIR spectra we identify for the first time in this source a bright, slightly
asymmetric HeI 10830 \AA~emission line (see Figure \ref{fig2}). This 
line was not reported in the NIR spectrum of \cite{rah08}, this detection may
imply variability in the wind of the source.  The asymmetry we detect may
indicate the presence of a P Cygni profile.

\section{\igrc}

\begin{figure}
\begin{tabular}{ccc}

\includegraphics[width=.4\textwidth,height=5truecm]{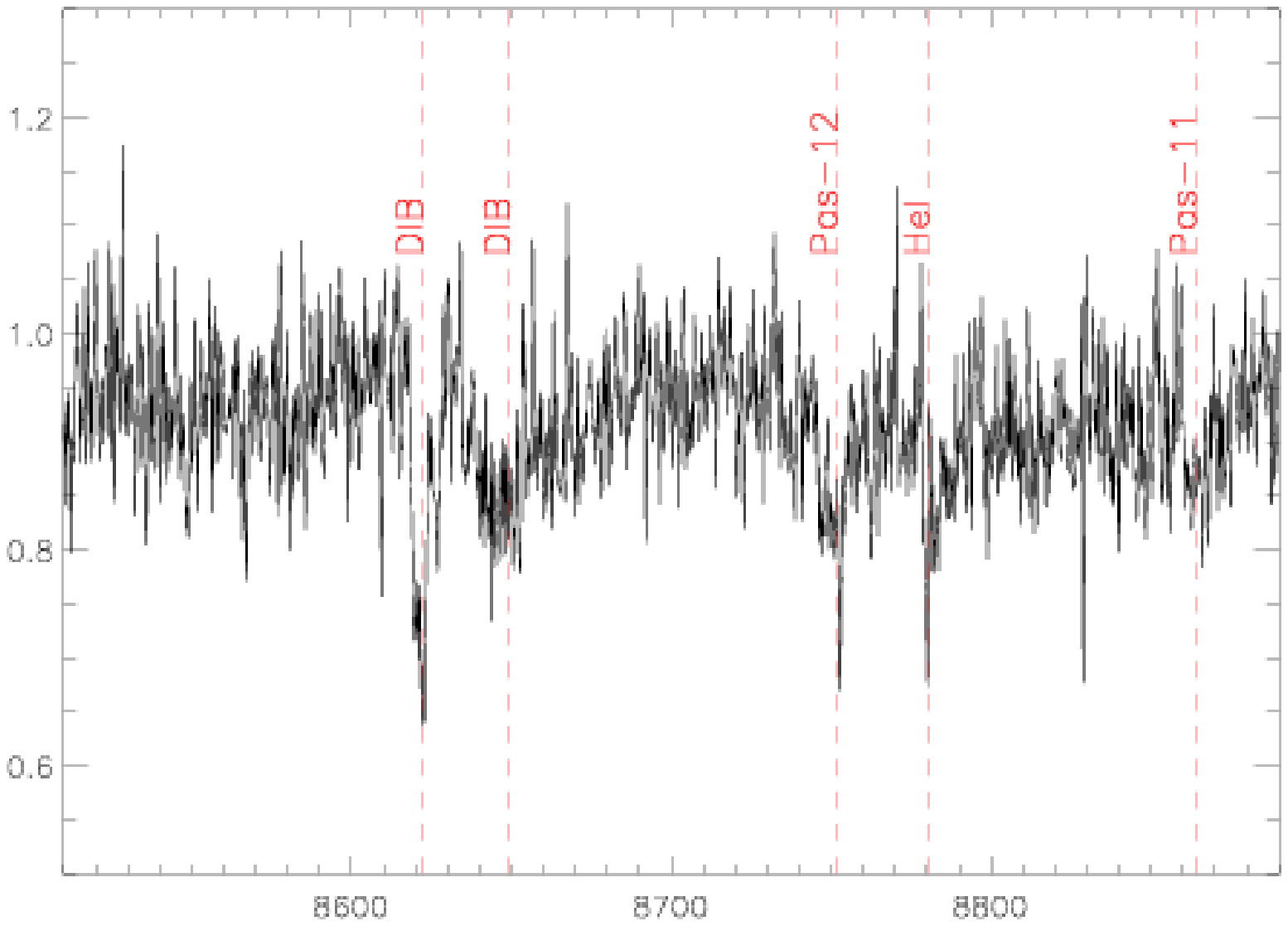} &
\includegraphics[width=.08\textwidth,height=5truecm]{white.ps} &
\includegraphics[width=.4\textwidth,height=5truecm]{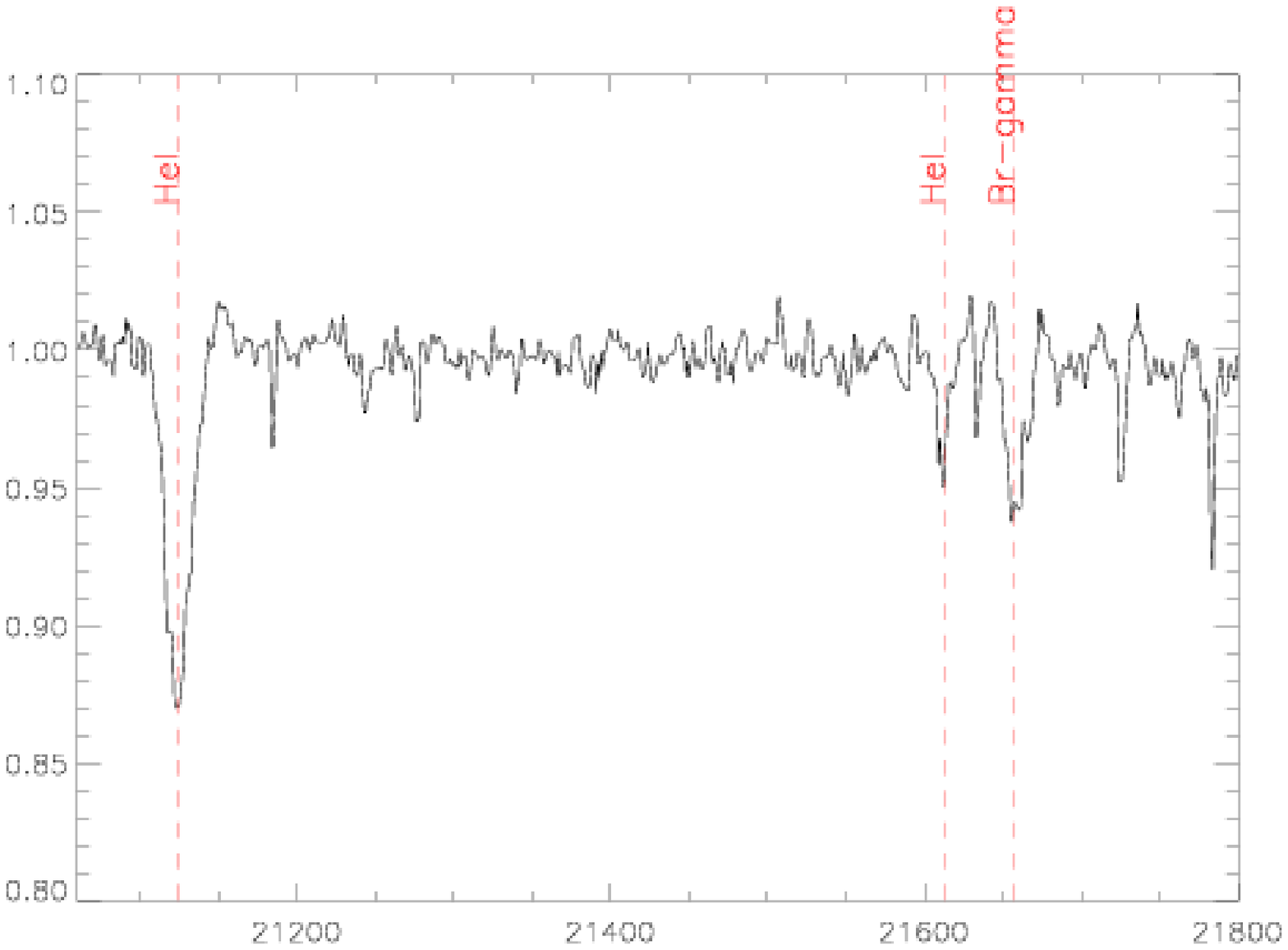} \\

\end{tabular}

\caption{Left: Section of the I band spectrum of \igrc. The low S/N allows only the detection of the strong Paschen lines and of
 diffuse interstellar bands.
Right: Section of the K band spectrum of \igrc~containing HeI 21126 \AA, He I 21614 \AA~ and Brackett $\gamma$ mainly  in absorption.
They also show small emission bumps consistent with the presence of
moderate winds. The absorption feature not marked between He I 21614 \AA~ and Brackett $\gamma$ is a telluric feature.}
\label{fig3}
\end{figure}

\igrc~ has been classified as an O8.5-O9.5 supergiant \cite{cha08} on the basis of photometric observations
and SED fitting but no spectroscopic observations have been reported yet. The source has been detected
with a low S/N (<10) in the VIS arm (Fig. \ref{fig3}). The Paschen lines 11 and 12 are visible and the continuum is  
quite faint. We therefore cannot use this part of the spectrum for a spectral type classification.
The 8620 \AA~DIB absorption has equivalent width $\sim$1.65
\AA~implying E(B-V)$\sim$4.49 and A(V)=13.9 for R(V)=3.1 consistent with \cite{rah08_1}.
Conversely the NIR spectrum has a good S/N ratio but it does not present strong
emission lines. We therefore try to establish a classification using
absorption features in the NIR spectrum. In the K band two common
features for O and B supergiants are the HeI 21126 \AA~
and the Brackett $\gamma$. Their equivalent widths are respectively
$\sim$ 2.8 \AA ~and  $\sim$1.2 \AA ~. According to \cite{han96}, these
EWs, especially the high value for HeI 21126 \AA, point
towards an O9.5 spectral type (see their figure 24).
These lines also display small emission excesses which may be due to
wind emission (Fig. \ref{fig3}).

\section{Conclusions}

We have presented the first results of our X-shooter observations of a sample of IGR sources.
Taking advantage of the unprecedented spectral coverage, we could compare the results in different
spectral domains. On this basis we suggest that \igra~is a B0.5-B1 supergiant star
in fast rotation, probably seen at high inclination. We confirm the
classification of \igrb~as a B0.5 supergiant
and we detect for the first time a bright He I emission line possibly suggesting enhanced mass loss
with respect to previous observations. We present the first VIS and NIR spectra of \igrc~ and on the basis
of the K band spectra we suggest that the star is a O9.5 supergiant.

Using the detection of the 8620 \AA~DIB in the three spectra we
estimate the reddening in the direction of these sources. The results
are compatible with published photometry when available. 
However in two of three cases the extinction estimations we obtain are compatible
with published ones only if we assume non standard extinction
curves.

\end{document}